# Dictionary Learning with Convolutional Structure for Seismic Data Denoising and Interpolation

Murad Almadani[*], Umair bin Waheed[†], Mudassir Masood[*], and Yangkang Chen[‡]

[*]*Department of Electrical Engineering*

*King Fahd University of Petroleum and Minerals*

*Dhahran, Saudi Arabia*

[†]*Department of Geosciences*

*King Fahd University of Petroleum and Minerals*

*Dhahran, Saudi Arabia*

[‡]*School of Earth Sciences*

*Zhejiang University*

*Hangzhou, China*

(November 8, 2024)

Running head: **Seismic data denoising and interpolation**

# ABSTRACT

Seismic data inevitably suffers from random noise and missing traces in field acquisition. This limits the utilization of seismic data for subsequent imaging or inversion applications. Recently, dictionary learning has gained remarkable success in seismic data denoising and interpolation. Variants of the patch-based learning technique, such as the K-SVD algorithm, have been shown to improve denoising and interpolation performance compared to the analytic transform-based methods. However, patch-based learning algorithms work on overlapping patches of data and do not take the full data into account during reconstruction. By contrast, the data patches (CSC) model treats signals globally and, therefore, has shown superior performance over patch-based methods in several image processing applications. In consequence, we test the use of CSC model for seismic data denoising and interpolation. In particular, we use the local block coordinate descent (LoBCoD) algorithm to reconstruct missing traces and clean seismic data from noisy input. The denoising and interpolation performance of the LoBCoD algorithm has been compared with that of K-SVD and orthogonal matching pursuit (OMP) algorithms using synthetic and field data examples. We use three quality measures to test the denoising accuracy: the peak signal-to-noise ratio (PSNR), the relative L2-norm of the error (RLNE), and the structural similarity index (SSIM). We find that LoBCoD performs better than K-SVD and OMP for all test cases in improving PSNR and SSIM, and in reducing RLNE. These observations suggest a huge potential of the CSC model in seismic data denoising and interpolation applications.

# INTRODUCTION

In seismic acquisition, the recorded data not only contain our signal of interest but also unwanted signals or noise. This contamination is caused by many sources including ocean waves, wind, instrument noise, and traffic, etc. To extract maximum value of the acquired data, noise attenuation is a crucial step in the seismic processing workflow. It enhances the signal quality, making it suitable for subsequent processing and interpretation. On the other hand, due to cost constraints and ground surface restrictions, in addition to regulatory reasons, we encounter missing traces that need to be interpolated. Many algorithms have been proposed addressing the issue of seismic data denoising and interpolation (Oropeza and Sacchi, 2011; Wang et al., 2015a; Tang and Ma, 2010; Bonar and Sacchi, 2012; Abma and Claerbout, 1995; Mousavi and Langston, 2016). Nevertheless, there is still much room for improvement in terms of algorithmic computational cost and reconstruction accuracy.

Sparse coding has been widely used as an effective tool to denoise and interpolate data, particularly in the field of image processing (Elad and Aharon, 2006; Zhao et al., 2009; Shen et al., 2009; Fadili et al., 2009; Turquais et al., 2017; Liu et al., 2018a). The underlying premise in sparse coding is that most signals and images in nature can be represented as a linear combination of a few elements from certain dictionary, such that it is distinguishable from the contaminated random noise. Also, it claims that the missing traces with zero values in the seismic data are present in a highly dense representation in the same dictionary. Therefore, finding the sparsest representation of data are the key to solving both problems.

Sparse transforms can be classified as analytic or learning based. Several analytic transforms have been proposed to denoise (Ibrahim and Sacchi, 2013; Neelamani et al., 2008; Kong and Peng, 2015; Kaplan et al., 2009) and interpolate (Gan et al., 2016; Wang et al., 2015b) seismic signals. This approach results in an implicit dictionary that is independent of the input data. However, searching for more representative dictionaries received a lot of attention to improve the denoising and interpolation performance.

Dictionary learning is one such approach where a dictionary is learned iteratively from the input data. Although learning a dictionary results in a higher computational complexity, it improves the denoising and interpolation performance significantly. During the past few years, dictionary learning has shown remarkable success in effectively denoising and interpolating 2-D seismic data, such as (Siahsar et al., 2017; Beckouche and Ma, 2014; Zhu et al., 2015; Chen et al., 2016; Zhang et al., 2018; Zu et al., 2019), whereas, (Chen, 2017; Nazari Siahsar et al., 2017) have leveraged this approach on higher dimensionality.

One of the most popular dictionary learning algorithms is the K-SVD algorithm (Aharon et al., 2006; Rubinstein et al., 2009), which it has been successfully applied in seismic data denoising and interpolation. However, K-SVD is a patch-based learning approach. This means that instead of taking into account the full image or signal, it works on overlapping patches. Therefore, it fails to detect global features which adversely affects the signal restoration accuracy (Romano and Elad, 2015).

Convolutional Sparse Coding (CSC) changes the patch-based learning approach by convolutional structure that treats the signals and images globally. This convolutional structure allows detection of the global signal and image features that cannot be detected using patch-based methods. In image processing, CSC has demonstrated superior performance in a variety of applications (Gu et al., 2015; Liu et al., 2016).

In CSC, seismic data are represented by a superposition of few small kernels convolved with sparse feature-maps. In general, CSC is a non-convex problem and the existing al-gorithms provide no guarantee to find its sparsest solution. However, converting the convolutional operation to element wise multiplication in frequency domain has recently been proposed making CSC more practical to use (Bristow et al., 2013; Bristow and Lucey, 2014; Kong and Fowlkes, 2014). This solution relies on the Alternating Direction Multipliers Minimization (ADMM) formulation (Boyd et al., 2011) by splitting the overall problem into subproblems, trying to find the sparse feature maps and the corresponding kernels in an iterative manner. ADMM formulation has recently been shown to successfully attenuate random and coherent noise in seismic data and in separating the ground roll (Liu et al., 2018b). However, the ADMM formulation suffers from the requirement of many tuning parameters that are application specific. More importantly, an inappropriate setting of these parameters could dramatically degrade the signal restoration performance, which in consequence, makes it harder to be implemented for various real-world problems.

Local block coordinate descent (LoBCoD) (Zisselman et al., 2018) algorithm avoids the ADMM formulation by converting the CSC model into global matrix-vector multiplication, resulting in a simplified implementation. The global dictionary matrix consists of shifted versions of local dictionaries in a circulant matrix form, and the sparse matrices are in an interlaced-concatenated vector form. Leveraging the concept of spark and mutual coherence of the dictionary matrix, Papyan et al. (2017) theoretically prove the uniqueness of the solution that could be obtained by using this CSC model and its stability in noisy cases. Moreover, the LoBCoD algorithm shows better performance than the previous CSC formulations (Bristow et al., 2013; Bristow and Lucey, 2014; Kong and Fowlkes, 2014) in overcoming the ADMM parameter tuning issue.

Here, we propose the use of the CSC model for seismic data denoising and interpolation. In particular, we leverage the LoBCoD algorithm to learn the dictionary and its correspond-ing global sparse vector. The results show superiority of LoBCoD over the analytical and patch-based techniques, such as the K-SVD algorithm. We validate these assertions through tests on both synthetic and field seismic data.

The rest of this paper is organized as follows. In the next section, we discuss the proposed LoBCoD algorithm for seismic denoising and interpolation along with a brief review of sparse coding theory. Numerical tests on synthetic and field data are presented afterwards.

## THEORY

### Sparse Coding Background

Conventionally, we represent the signal $\mathbf{x} \in \mathbb{R}^N$ by $\mathbf{x} = \mathbf{y} + \boldsymbol{\varepsilon}$, where $\mathbf{y} \in \mathbb{R}^N$ is an uncontaminated signal, and $\boldsymbol{\varepsilon} \in \mathbb{R}^N$ is an additive noise vector. However, in sparse coding, the signal $\mathbf{x}$ is represented through the following equation:

$$\mathbf{x} = \mathbf{D}\mathbf{z} + \boldsymbol{\varepsilon}, \tag{1}$$

where $\mathbf{D} \in \mathbb{R}^{N \times M}$ (M>N) is a dictionary matrix formed of $M$ columns called atoms or dictionary elements. Sparse coding learns the dictionary $\mathbf{D}$ and $M$-dimensional vector $\mathbf{z}$ such that the denoised version of signal $\mathbf{x}$ can be approximated by $\hat{\mathbf{x}} = \mathbf{D}\mathbf{z}$. $\mathbf{z}$ is a sparse vector which implies that it has $m << M$ non-zero coefficients. Thus, we aim to reconstruct $\hat{\mathbf{x}}$ from only a few dictionary elements. Choosing an effective dictionary to represent the signal sparsely is the most crucial aspect of this model to succeed. This dictionary could be a wavelet, curvelet (Candès and Donoho, 2004), bandelet (Le Pennec and Mallat, 2005), or a contourlet (Do and Vetterli, 2003) dictionary, as these domains are able to sparsify most signals and images. Many algorithms have been proposed (e.g., Orthogonal Matching

Pursuit (Chen et al., 1989), Basis Pursuit (Chen et al., 2001)) to find the unique sparsest solution of $\mathbf{z}$ that preserves some percentage of data consistency by solving the minimization problem

$$\mathbf{z} = \arg\min_{\mathbf{z}} \|\mathbf{x} - \mathbf{D}\mathbf{z}\|_2^2 + \lambda\|\mathbf{z}\|_1, \qquad (2)$$

where $\lambda$ is a regularization parameter. Although the dictionaries mentioned above provide sparse representations of most signals and images in nature, the sparsity could be improved further by using a dictionary learned from the available data. Nevertheless, this approach of dictionary learning cannot be used for images because of prohibitive computational complexity owing to their size. Therefore, a workaround is to perform dictionary learning with the help of patches extracted from the image itself. Consequently, sparse coding solves the minimization problem to learn the dictionary $\mathbf{D}$ and the sparse vector $\mathbf{z}$ through

$$\min_{\mathbf{D},\mathbf{z}} \sum_{ij} \|R_{ij}(\mathbf{x}) - \mathbf{D}\mathbf{z}_{ij}\|_2^2 \quad s.t. \ \|\mathbf{d}_k\|_2 = 1 \ \forall k, \ \ \|\mathbf{z}_{ij}\|_0 < T_0 \ \forall i,j, \qquad (3)$$

where $\mathbf{D}$ is the learned dictionary that contains unit-norm columns $\mathbf{d}_k$, such that, $1 \leq k \leq M$. $R_{ij}$ is the patches extraction operator, and $\mathbf{z}_{ij}$ is the sparse representation of each image patch that is restricted to have a number of non-zero elements less than $T_0$ i.e., $\|\mathbf{x}\|_0 < T_0$. In this regard, along with other algorithms, method of optimal directions (MOD) (Engan et al., 1999) and K-SVD (Aharon et al., 2006) have been proposed for solving this optimization task and have shown better signal and image restoration performance.

However, since this remedy relies on image patches, it fails to utilize global image features which adversely affects the signal restoration accuracy. Convolutional sparse coding (CSC) changes the patch-based learning approach by convolutional structure that treats the signals and images globally.

## LoBCoD for Seismic Data Denoising

The CSC model represents a signal $\mathbf{x} \in \mathbb{R}^N$ as a sum of $m$ convolutions, built by feature maps $[\mathbf{z}_k]_{k=1}^m$, convolved with $m$ local filters $[\mathbf{d}_k]_{k=1}^m$. These filters are of length $n << N$. The dictionary learning problem can then be formulated as the following minimization problem over the filters and the feature maps:

$$\min_{\mathbf{d},\mathbf{z}} \frac{1}{2}\|\mathbf{x} - \sum_{k=1}^m \mathbf{d}_k * \mathbf{z}_k\|_2^2 + \lambda \sum_{k=1}^m \|\mathbf{z}_k\|_1 \quad s.t. \ \|\mathbf{d}_k\|_2 \leq 1, \tag{4}$$

where the symbol $*$ denotes the convolution operator.

This convolutional formulation of the dictionary learning problem has the advantage of using global image features while learning dictionary. However, the resulting optimization problem is non-convex and the existing algorithms provide no guarantees to find its optimum solution. ADMM (Boyd et al., 2011) has been leveraged to solve this dilemma by converting the convolutional operator to a multiplication in frequency domain. However, this approach suffers from its many tuning parameters. These parameters are application specific and therefore great care has to be taken when setting them. Inappropriate settings of these parameters could dramatically degrade signal reconstruction performance.

The LoBCoD algorithm avoids the need to use the ADMM formulation by converting the CSC model into a global matrix-vector multiplication. This converts the problem in equation 4 to:

$$\min_{\mathbf{D},\mathbf{\Gamma}} \frac{1}{2}\|\mathbf{x} - \mathbf{D}\mathbf{\Gamma}\|_2^2 + \lambda\|\mathbf{\Gamma}\|_1, \tag{5}$$

where $\mathbf{D}$ is the global dictionary composed of all shifted versions of a local dictionary $\mathbf{D}_L$ of size $n \times m$, and $\mathbf{\Gamma}$ is the global sparse vector, which is the interlaced concatenation of all feature maps $[\mathbf{z}_k]_{k=1}^m$, as illustrated in Figure 1.

[Figure 1 about here.]

Finding the sparsest solution of $\boldsymbol{\Gamma}$ globally is not feasible due to its huge dimension; therefore, splitting the global optimization problem to finding small sparse vectors is the core idea of solving this problem.

Consequently, the global sparse vector $\Gamma$ is split into $N$ non-overlapping $m$-dimensional vectors $\boldsymbol{\alpha}_k$, also known as needles, such that, each of these needles operates on a single local dictionary $\mathbf{D}_L$ in the global dictionary matrix $\mathbf{D}$. Using this, the noisy signal $\mathbf{x}$ can be expressed as:

$$\mathbf{x} = \sum_{k=1}^{N} \mathbf{P}_k^T \mathbf{D}_L \boldsymbol{\alpha}_k, \qquad (6)$$

where $\mathrm{P}_k^T \in \mathbb{R}^{N \times n}$ is an operator that localizes $\mathbf{D}_L \boldsymbol{\alpha}_k$ into the $k$-th position of a vector of length $N$ and pads the remaining vector values by zeros. This converts equation 5 into:

$$\min_{\boldsymbol{\alpha}_i} \frac{1}{2} \left\| \mathbf{x} - \sum_{k=1}^{N} \mathbf{P}_k^T \mathbf{D}_L \boldsymbol{\alpha}_k \right\|_2^2 + \lambda \sum_{k=1}^{N} \|\boldsymbol{\alpha}_k\|_1, \qquad (7)$$

which finds the sparse feature maps over a fixed dictionary. Note that the dictionary learning step is skipped in equation 7 as the sparse coding and dictionary learning steps take place in an alternating fashion.

Instead of optimizing with respect to all the needles together, the LoBCoD algorithm optimizes each needle $\boldsymbol{\alpha}_k$ separately. Therefore, the update rule for each needle can be written as:

$$\min_{\boldsymbol{\alpha}_k} \frac{1}{2} \left\| (\mathbf{x} - \sum_{\substack{j=1 \\ j \neq k}}^{N} \mathbf{P}_j^T \mathbf{D}_L \boldsymbol{\alpha}_j) - \mathbf{P}_k^T \mathbf{D}_L \boldsymbol{\alpha}_k \right\|_2^2 + \lambda \|\boldsymbol{\alpha}_k\|_1. \qquad (8)$$

To simplify notation, we define

$$\mathbf{R}_k = (\mathbf{x} - \sum_{\substack{j=1 \\ j \neq k}}^{N} \mathbf{P}_j^T \mathbf{D}_L \boldsymbol{\alpha}_j), \qquad (9)$$

as the residual signal without the contribution from the $k$-th needle $\boldsymbol{\alpha}_k$. The above equation 8 can then be re-written as:

$$\min_{\boldsymbol{\alpha}_k} \frac{1}{2} \|\mathbf{R}_k - \mathbf{P}_k^T \mathbf{D}_L \boldsymbol{\alpha}_k\|_2^2 + \lambda \|\boldsymbol{\alpha}_k\|_1. \tag{10}$$

The above minimization can be decomposed into an equivalent and local problem ( see Appendix A of Zisselman et al. (2018)):

$$\min_{\boldsymbol{\alpha}_k} \frac{1}{2} \|\mathbf{P}_k \mathbf{R}_k - \mathbf{D}_L \boldsymbol{\alpha}_k\|_2^2 + \lambda \|\boldsymbol{\alpha}_k\|_1. \tag{11}$$

After estimating the sparse feature maps, the global dictionary matrix update is done by updating its local matrix $\mathbf{D}_L$. This is performed by solving the following minimization problem, subject to the constraint of normalized dictionary columns:

$$\min_{\mathbf{D}_L} \frac{1}{2} \left\| \mathbf{x} - \sum_{k=1}^{N} \mathbf{P}_k^T \mathbf{D}_L \boldsymbol{\alpha}_k \right\|_2^2 \quad s.t. \ \|[\mathbf{d}_k]_{k=1}^m\|_2 = 1. \tag{12}$$

This can be achieved by using projected steepest descent. The gradient of the quadratic term in equation 12 w.r.t. $\mathbf{D}_L$ is:

$$\nabla \mathbf{D}_L = -\sum_{k=1}^{N} \mathbf{P}_k (\mathbf{x} - \hat{\mathbf{x}}) \boldsymbol{\alpha}_k^T, \tag{13}$$

where $\hat{\mathbf{x}}$ is the most recent reconstructed signal. The final update step for $\mathbf{D}_L$ is given as:

$$\mathbf{D}_L = \mathcal{P}[\mathbf{D}_L - \eta \nabla \mathbf{D}_L], \tag{14}$$

where $\eta$ is the step size, and $\mathcal{P}$ is a normalization operator that forces dictionary columns $[\mathbf{d}_k]_{k=1}^m$ to have unit norm. It is pertinent to note that updating $\mathbf{D}_L$ after updating the entire global sparse vector $\boldsymbol{\Gamma}$ is inefficient and leads to extremely slow convergence, particularly when $N$ is large. Therefore, we use the stochastic gradient descent approach to compute the gradient of $\mathbf{D}_L$ after updating only a small number of sparse needles at a time in a stochastic manner. In this regard, instead of concluding the entire pursuit stage and then progressing towards the global minimum in equation 14, we proceed by taking a small step size $\eta$ after estimating a small number of needles $\boldsymbol{\alpha}$ (Bottou and Bousquet, 2008).

After iterative updates of the global dictionary $\mathbf{D}$ and its global sparse vector $\boldsymbol{\Gamma}$, the denoised signal is reconstructed as:

$$\hat{\mathbf{x}} = \sum_{k=1}^{N} \mathbf{P}_k^T \mathbf{D}_L \boldsymbol{\alpha}_k. \tag{15}$$

In case of perfect denoising, the reconstructed signal $\hat{\mathbf{x}}$ should equal the desired clean signal $\mathbf{y}$. The proposed denoising algorithm is summarized in Algorithm 1.

---
**Algorithm 1:** LoBCoD based denoising algorithm summary

**Input**: Vectorized noisy seismic data ($\mathbf{x}$), initial local dictionary ($\mathbf{D}_L$)

**Main iteration**:

1) Calculate all the sparse needles $\boldsymbol{\alpha}_k$ via equation 11

2) Update the local dictionary $\mathbf{D}_L$ via equations 13 and 14

**Output**: Denoised seismic data ($\hat{\mathbf{x}} = \sum_{k=1}^{N} \mathbf{P}_k^T \mathbf{D}_L \boldsymbol{\alpha}_k$)

---

## LoBCoD for Seismic Data Interpolation

The core task of seismic interpolation is to accurately fill the missing values/traces in seismic data as if they have been recorded via sensors. In sparse coding, these missing traces are represented by dense feature maps using good representative dictionaries of the desired seismic data (Siahsar et al., 2017). Therefore, estimating the sparsest feature maps will provide interpolated traces that harmonize with the already detected ones. These good representative dictionaries could be pre-trained via complete seismic data, or trained directly from the corrupted seismic data itself. This can be done by solving the following

optimization task:

$$\min_{\mathbf{D},\mathbf{\Gamma}} \frac{1}{2}\|\mathbf{x} - \mathrm{A}\mathbf{D}\mathbf{\Gamma}\|_2^2 + \lambda\|\mathbf{\Gamma}\|_1, \qquad (16)$$

where A is the mask operator with the same size of the corrupted seismic data $\mathbf{x}$, such that it indicates the location of missing traces with zero values; consequently, $\mathbf{x} = \mathrm{A}\mathbf{y}$, such that, $\mathbf{y}$ is the desired full sampled data.

Leveraging the concept used in the denoising part of defining the sparse needles $\boldsymbol{\alpha}_k$, and by taking into account the mask operator A, the problem is converted as follows:

$$\min_{\boldsymbol{\alpha}_k} \frac{1}{2}\|\mathbf{P}_k\mathbf{R}_k - \mathrm{A}_k\mathbf{D}_L\boldsymbol{\alpha}_k\|_2^2 + \lambda\|\boldsymbol{\alpha}_k\|_1, \qquad (17)$$

where $\mathrm{A}_k$ is the operator that masks the $k$-th location of the global sparse vector $\mathbf{\Gamma}$, and it is equal to $\mathbf{P}_k\mathrm{A}\mathbf{P}_k^T$, whereas,

$$\mathbf{R}_k = (\mathbf{x} - \mathrm{A}_k \sum_{\substack{j=1\\j\neq k}}^{N} \mathbf{P}_j\mathbf{D}_L\boldsymbol{\alpha}_k). \qquad (18)$$

As has been discussed in the previous section on denoising, the final solution is obtained by performing a number of iterations between updating the sparse needles $\boldsymbol{\alpha}_k$ and the local dictionary $\mathbf{D}_L$. In this regard, the local dictionary $\mathbf{D}_L$ is updated via the gradient descent, as follows:

$$\nabla \mathbf{D}_L = -\sum_{k=1}^{N} \mathbf{P}_k \mathrm{A}^T (\mathbf{x} - \mathrm{A}\hat{\mathbf{x}})\boldsymbol{\alpha}_k^T, \qquad (19)$$

such that the final reconstructed seismic data are given by:

$$\hat{\mathbf{x}} = \sum_{k=1}^{N} \mathbf{P}_k^T \mathbf{D}_L \boldsymbol{\alpha}_k. \qquad (20)$$

The proposed LoBCoD based interpolation algorithm is summarized in Algorithm 2.

**Algorithm 2:** LoBCoD based interpolation algorithm summary

**Input**: Vectorized seismic data with missing traces ($\mathbf{x}$), initial local dictionary ($\mathbf{D}_L$)

**Main iteration**:

1) Calculate all the sparse needles $\boldsymbol{\alpha}_k$ via equation 17

2) Update the local dictionary $\mathbf{D}_L$ via equations 19 and 14

**Output**: Interpolated seismic data ($\hat{\mathbf{x}} = \sum_{k=1}^{N} \mathbf{P}_k^T \mathbf{D}_L \boldsymbol{\alpha}_k$)

## NUMERICAL TESTS

In this section, we test performance of the LoBCoD algorithm in denoising and interpolating seismic data. We compare its performance with that of the K-SVD and the OMP algorithms (Pati et al., 1993). For the LoBCoD algorithm, we use 100 filters of size 11×11 with a step size of $\eta$=0.1. On the other hand, for the K-SVD algorithm, we set the patch size to 11×11 and the dictionary to have 512 atoms with 30% of image patches used in dictionary learning process. While, we used DCT dictionary with 512 atoms in OMP. Also, we use a step size equal to 1 when retrieving the patches in OMP and K-SVD. Figure 2 shows some atoms/kernels learned from the field seismic data of Figure 11 using both K-SVD and LoBCoD algorithms. The elements of discrete cosine transform dictionary (DCT) used in the OMP algorithm are also shown in this figure.

[Figure 2 about here.]

We use the peak signal-to-noise ratio (PSNR) as a metric to measure the denoising performance. PSNR is defined as:

$$\text{PSNR} = 20 \log_{10} \frac{\mathbf{x}_{max}}{\|\mathbf{x} - \hat{\mathbf{x}}\|_2^2}, \tag{21}$$

where **x** is the original signal before noise addition, $\hat{\mathbf{x}}$ is the reconstructed signal, and $\mathbf{x}_{max}$ is the maximum possible value of the data. In our case, $\mathbf{x}_{max} = 1$ since we normalize each trace before denoising.

We also use the relative L2-norm of the error (RLNE) (Qu et al., 2012, 2014) as a measure to quantify closeness of the reconstructed signal to the original signal along with the structural similarity index (SSIM) (Bovik et al., 2004). RLNE is defined as:

$$\text{RLNE} = \frac{\|\hat{\mathbf{x}} - \mathbf{x}\|_2}{\|\mathbf{x}\|_2}, \tag{22}$$

where x and $\hat{\text{x}}$ are defined as above. An RLNE value of zero indicates perfect reconstruction, while errors in reconstruction result in larger RLNE values.

SSIM indicates the structural similarity of the estimated data with respect to the ground truth. It calculates the ability of the reconstruction algorithm to preserve details of the original data. SSIM is defined as:

$$\text{SSIM} = \frac{(2\mu_x \mu_{\hat{x}} + c_1)(2\sigma_{x\hat{x}} + c_2)}{(\mu_x^2 + \mu_{\hat{x}}^2 + c_1) + (\sigma_x^2 + \sigma_{\hat{x}}^2 + c_2)}, \tag{23}$$

where $\mu_x, \mu_{\hat{x}}, \sigma_x^2, \sigma_{\hat{x}}^2$, and $\sigma_{x\hat{x}}$ is the mean, variance, and covariance of the ground truth image **x** and the reconstructed image $\hat{\text{x}}$, whereas, $c_1$ and $c_2$ are constants related to the dynamic range of the data. SSIM values range between [0,1]; a higher value indicates better reconstruction performance.

## Synthetic Data Denoising

In this section, we consider a synthetic shot record for the BP 2004 velocity model (Billette and Brandsberg-Dahl, 2005) and add random noise corresponding to input PSNR values of 20 dB and 15 dB. We measure improvement in the PSNR value for each case due to denoising using the OMP, K-SVD, and LoBCoD algorithms. We also monitor deterioration in reconstruction accuracy using the RLNE and SSIM measures for all algorithms as the noise level rises.

[Figure 3 about here.]

Figure 3 shows the shot record used for denoising tests. We add white Gaussian noise to this record before feeding it to the denoising algorithms. Figure 4a shows noisy data with a PSNR value of 20 dB. Figure 4b-4d show denoised signals using the OMP, K-SVD, and LoBCoD algorithms, respectively. The reconstruction using the OMP and K-SVD algorithms improves the PSNR value to 32.38 dB and 33.43 dB, respectively, while LoBCoD improves the PSNR to 34.31 dB. The input noisy signal (Figure 4a) has an RLNE value of 1.48, whereas the OMP and the K-SVD based denoising results in RLNE values of 0.35 and 0.32, respectively, and the LoBCoD algorithm reduces the RLNE value to 0.29. For both quality metrics, we find that the LoBCoD algorithm performs slightly better than OMP and K-SVD.

[Figure 4 about here.]

Next, we increase the noise level to observe how reconstruction accuracy of these algorithms is affected by the rising noise level in the input data. Figure 5 shows the noisy data with a PSNR value of 15 dB (Figure 5a), the reconstructed data using OMP (Figure 5b), K-SVD (Figure 5c), and the one using LoBCoD (Figure 5d). OMP, K-SVD and LoBCoD based denoising improve the PSNR to 29.21 dB, 30.34 dB, and 31.80 dB, respectively. The RLNE value for the input noisy signal, in this case, is 2.64 which improves to 0.51, 0.45, and 0.38 for OMP, K-SVD, and LoBCoD, respectively. In this case, we observe relatively larger improvement in signal reconstruction using the LoBCoD algorithm as compared to the OMP and K-SVD algorithms.

[Figure 5 about here.]

[Table 1 about here.]

[Table 2 about here.]

Furthermore, we calculate the structural similarity index (SSIM) of the reconstructed data to test the ability of the considered denoising algorithms in preserving fine details of the ground truth seismic data (see Table 3). The results show superiority of the LoBCoD algorithm in preserving structural details of the original data for all noise levels. Remember, an SSIM value close to 1 indicates better performance.

Although LoBCoD performs better than OMP and K-SVD in each test case, most notable is the fact that as the noise level in the input signal rises, the quality of reconstruction for LoBCoD does not deteriorate as fast as that for K-SVD. This is because K-SVD is a patch-based technique and, therefore, it does not take global features into account during reconstruction. This results in a compromised performance when the noise level is high, as observed in Figure 6 and in the residual errors which appear in Figures 7 and 8. On the contrary, LoBCoD treats the signal globally. Thus, in noisy conditions the reconstruction accuracy of LoBCoD does not suffer as much as that of K-SVD. These observation can also be seen in the zoomed displays shown in Figures 9 and 10. Moreover, we find the CSC based reconstruction is computationally slightly faster than the patch-based method, as it takes 173 s and 216 s for both approaches to converge, respectively. We performed this experiment on a computer with Intel Core i7 CPU at 2.67 GHz and 8GB RAM, and with data size 2001 × 1201. This observation is consistent with that made by Quan and Jeong (2016).

[Table 3 about here.]

[Figure 6 about here.]

[Figure 7 about here.]

[Figure 8 about here.]

[Figure 9 about here.]

[Figure 10 about here.]

**Field Data Denoising**

We consider a shot gather from the Nile Delta, Egypt as shown in Figure 11. It is an unprocessed seismic data and contains useful signal with contaminated random noise. The data cover an offset of 15 km with 12 s of recording. Note that, the first breaks occur around 2 s, thus the signal before that time could be treated as noise. This can help in evaluating the denoising algorithms as there is no ground truth signal to compare with.

In this experiment, we denoise the data in Figure 11 using the OMP, K-SVD, and the proposed LoBCoD approach. The resulting denoised images are shown in Figures 12a-12c, respectively. A close observation of the figures indicate the superior performance of the LoBCoD approach in removing the noise. However, we quantitatively compare the performance of different methods by a numerical test. Specifically, we calculate the average energy ratio of the signals before and after 2 s. This ratio is indicative of how much of the useful energy has been preserved with respect to the filtered noise during the reconstruction. In Table 4, we refer to the calculated average energy of the signals before and after 2 s by S1 and S2, respectively. The results show that LoBCoD yields the highest S2/S1 ratio with 102.06 compared to 99.57 and 89.53 for the K-SVD and OMP, respectively. This indicates that LoBCoD exhibits superiority in preserving useful signals while better filtering out the noise.

[Table 4 about here.]

We also calculate the residual error for all the approaches with respect to the original noisy data of Figure 11. The residual errors are displayed in Figure 13 and show the superior performance of LoBCoD in filtering out the noise without destroying the useful signal.

[Figure 11 about here.]

[Figure 12 about here.]

[Figure 13 about here.]

To support our claim, we plot in Figure 14 the zoom-in displays of a section of the seismic events and the corresponding residual errors for all three approaches. The region located between 8-12 s and 0-7 km offset was selected for this purpose. It is obvious from these figures that OMP failed to remove a significant amount of noise as compared to both K-SVD and LoBCoD. In addition, a structure is clearly visible in the residual error plot of K-SVD (Figure 14d). This indicates that a part of useful seismic events has been filtered out along with the noise by the K-SVD approach. Such inconsistencies in denoising cannot be observed in LoBCoD which further demonstrate the robustness of the proposed approach.

[Figure 14 about here.]

## Interpolation of a Field Seismic Data With Missing Traces

[Figure 15 about here.]

We also examine the performance of LoBCoD algorithm in interpolating missing traces in seismic data and compare it with the K-SVD and OMP interpolation algorithms. In this regard, we use a three-dimensional seismic data taken from the Teapot Dome survey (Oren and Nowack, 2018) with $512 \times 256 \times 64$ grid point, as shown in Figure 15. We randomly zero out 30%, 40%, and 50% of the seismic traces and add random Gaussian noise corresponding to input PSNR value of 20 dB, as shown in Figures 16a, 17a, and 18a, respectively. Figures 16b, 17b, and 18b show the corresponding interpolation results using OMP algorithm, whereas, the corresponding K-SVD interpolation algorithm results are shown in Figures 16c, 17c, and 18c, respectively. Likewise, LoBCoD interpolation algorithm results are shown in Figures 16d, 17d, and 18d.

The results show superior interpolation performance of the LoBCoD algorithm compared to K-SVD and OMP for all cases. For a quantitative assessment, we calculate PSNR, RLNE, and SSIM values for the interpolation results with respect to the ground truth data shown in Figure 15. The results are summarized in Tables 5-7 for the 30%, 40%, and 50% missing traces cases, respectively. We observe that LoBCoD exhibits the best performance. We would especially like to highlight that as the number of missing traces is increased, the performance of the proposed approach does not deteriorate as fast as the performance of the other algorithms, as it is seen in the residual error plots of the 50% missing traces case shown in Figure 19. Moreover, we observe artifacts in the reconstructed data cubes for OMP and K-SVD that increase rapidly as we increase the number of missing traces. However, we do not observe this with LoBCoD based reconstruction as the problem starts to show up only on a few traces when 50% of the data are missing. This explains the importance of treating the data globally while learning its dictionary during the reconstruction. As we have stated earlier, K-SVD (unlike LoBCoD) is a patch-based learning approach which works on overlapping patches of the data and do not take the full data textures into account during dictionary learning, which results in a sub-optimal dictionary. This dictionary along with the dictionary used in OMP (the DCT) fail to sparsify the data very well, and therefore, has a deteriorated performance as compared to LoBCoD.

[Figure 16 about here.]

[Figure 17 about here.]

[Figure 18 about here.]

[Figure 19 about here.]

[Table 5 about here.]

[Table 6 about here.]

[Table 7 about here.]

# CONCLUSION

We studied the problem of seismic data denoising and interpolation using the local block coordinate descent (LoBCoD) algorithm and compared it with the OMP and K-SVD algorithms. LoBCoD implements the convolution sparse coding (CSC) model for dictionary learning while the OMP and K-SVD are analytical and patch-based dictionary learning methods, respectively. We use three metrics for testing the quality of reconstructed data, namely the peak signal-to-noise-ratio (PSNR), the relative L2-norm of the error (RLNE), and the structural similarity index (SSIM). We find that, unlike OMP and K-SVD, the LoBCoD algorithm exhibits superior performance in denoising and interpolating synthetic and field seismic data. This superior performance is due to the capability of CSC to capture and utilize global features. We deduce this by observing improvement in PSNR and SSIM values, along with reduction in RLNE value for all test cases, particularly when the noise level is high. Also, we observed the implementation simplicity of the proposed LoBCoD algorithm compared with other CSC techniques, e.g. ADMM algorithm (Boyd et al., 2011). This simplicity comes from the fact that it requires only one regularisation parameter, which depends on the detected noise level.

# LIST OF FIGURES

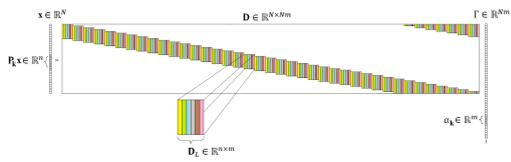

Figure 1: Matrix-vector representation of CSC model used in the LoBCoD algorithm (Zisselman et al., 2018).

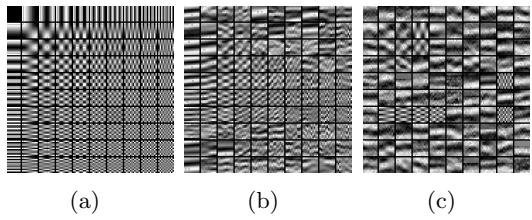

Figure 2: (a) DCT dictionary used in the OMP algorithm. (b) 100 dictionary atoms learned using the K-SVD algorithm. (c) 100 kernels learned using the LoBCoD algorithm.

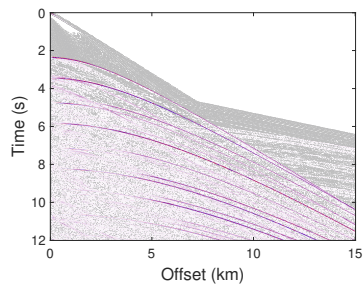

Figure 3: Synthetic shot record for the BP 2004 velocity model used for denoising tests.

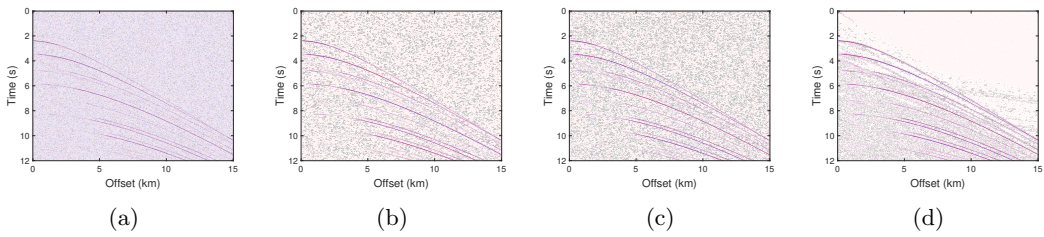

Figure 4: Input noisy data with a PSNR of 20 dB (a), and the reconstructed data using the OMP (b), the K-SVD (c), and the LoBCoD (d) algorithm.

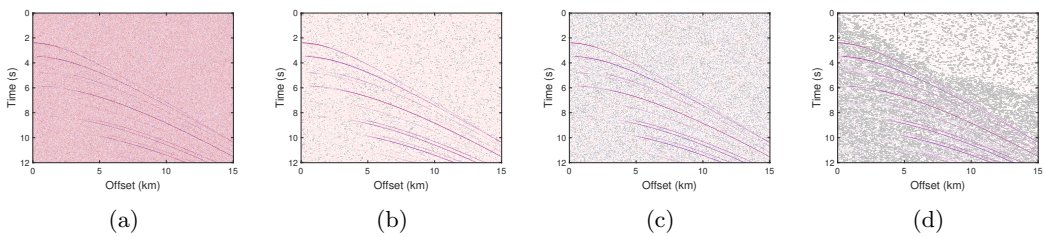

Figure 5: Input noisy data with a PSNR of 15 dB (a), and the reconstructed data using the OMP (b), the K-SVD (c), and the LoBCoD (d) algorithm.

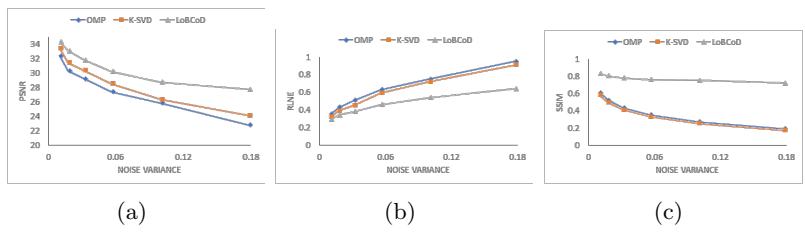

| (a) | (b) | (c) |

Figure 6: Denoising performance of the OMP, K-SVD, and LoBCoD algorithms as noise variance increases with respect to PSNR (a), RLNE (b), and SSIM (c) values.

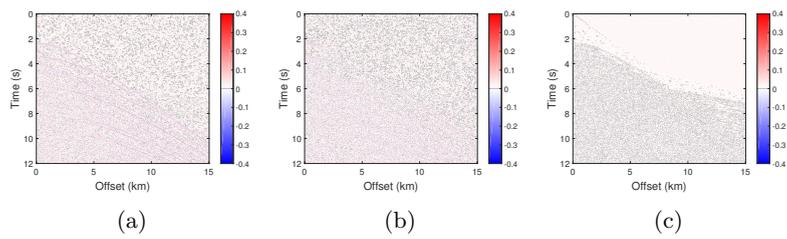

Figure 7: Denoising errors using (a) OMP, (b) K-SVD, and (c) LoBCoD algorithm for the results shown in Figure 4.

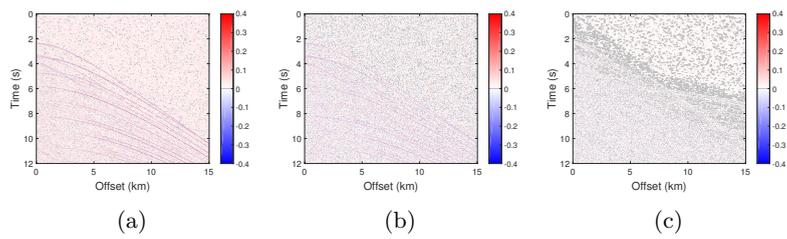

Figure 8: Denoising errors using (a) OMP, (b) K-SVD, and (c) LoBCoD algorithm for the results shown in Figure 5.

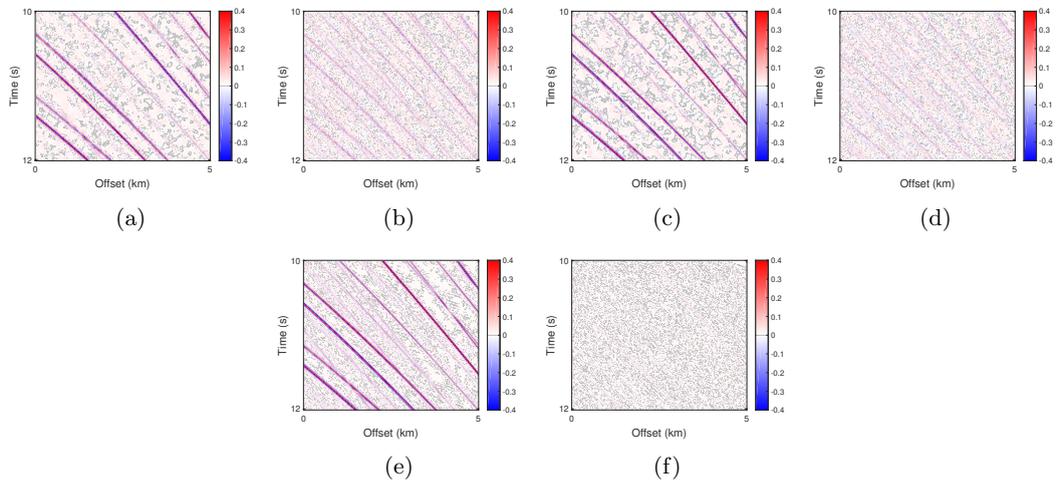

Figure 9: Zoom-in display of the seismic events occurring at 10-12 s and 0-5 km offset. (a), (c), and (e) are the zoomed displays of the reconstructed data shown in Figures 4b, 4c, and 4d, respectively. (b), (d), and (f) are the zoomed displays of the residual errors shown in Figure 7.

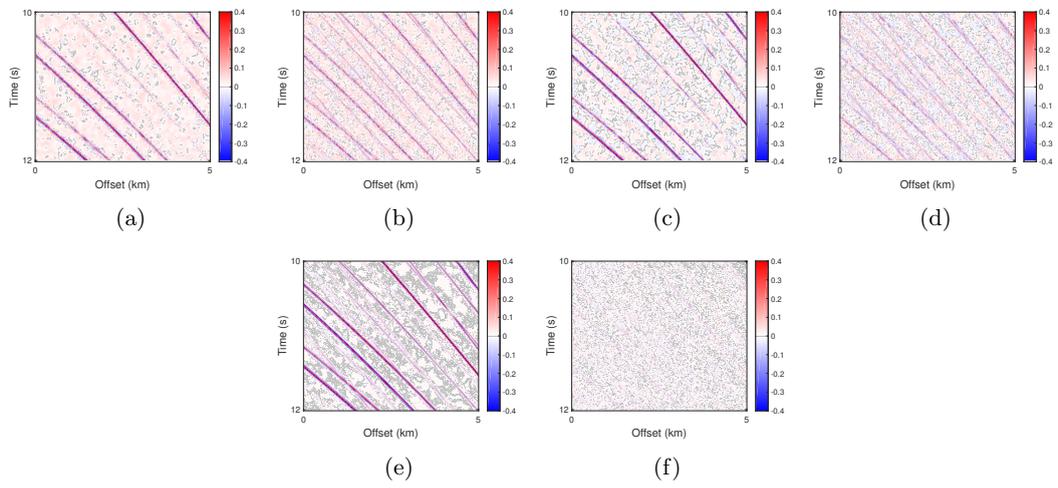

Figure 10: Zoom-in display of the seismic events occurring at 10-12 second and 0-5 km offset. (a), (c), and (e) are the zoomed displays of the reconstructed data shown in Figures 5b, 5c, and 5d, respectively. (b), (d), and (f) are the zoomed displays of the residual errors shown in Figures 8.

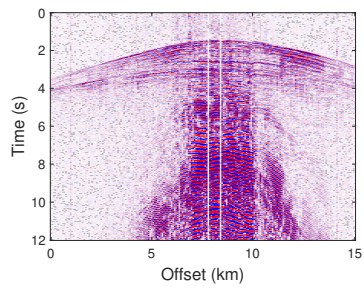

Figure 11: Noisy field seismic data taken from the Nile Delta, Egypt.

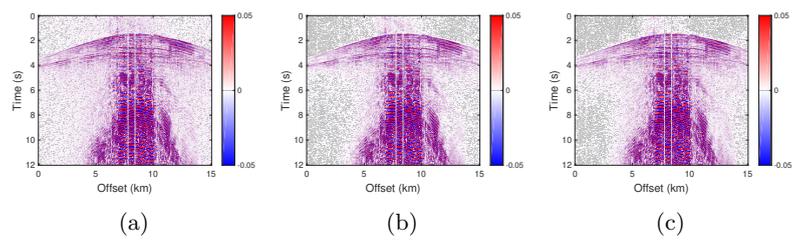

Figure 12: Field data denoising using the (a) OMP, (b) K-SVD, and (c) LoBCoD algorithm.

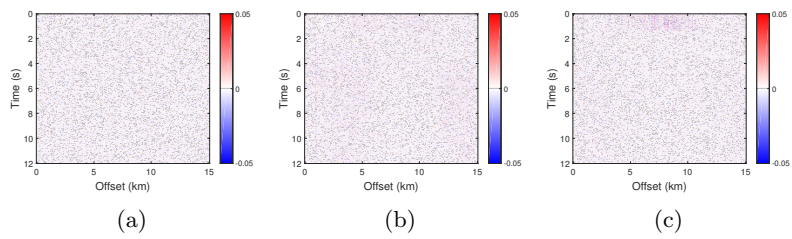

Figure 13: Estimated errors using (a) OMP, (b) K-SVD, and (c) LoBCoD algorithm for the field seismic case results shown in Figure 12.

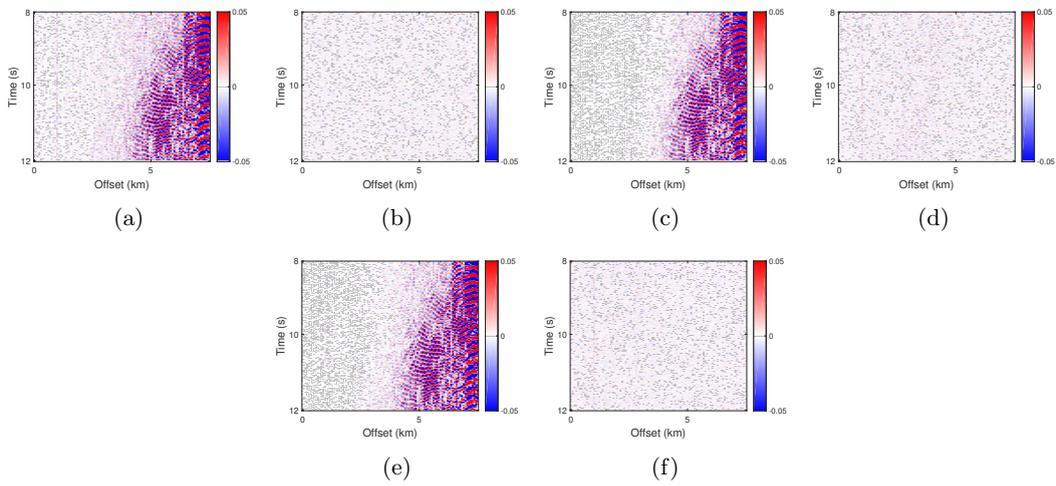

Figure 14: Zoom-in display of the seismic events occurring at 8-12 s and 0-7 km offset. (a), (c), and (e) are the zoomed displays of the reconstructed data shown in Figure 12. (b), (d), and (f) are the zoomed displays of the estimated noise shown in Figure 13.

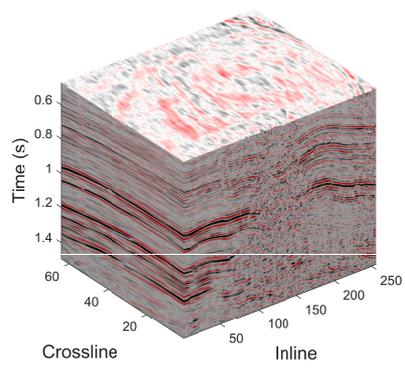

Figure 15: 3D seismic data taken form the Teapot Dome survey (Oren and Nowack, 2018) used for interpolation tests.

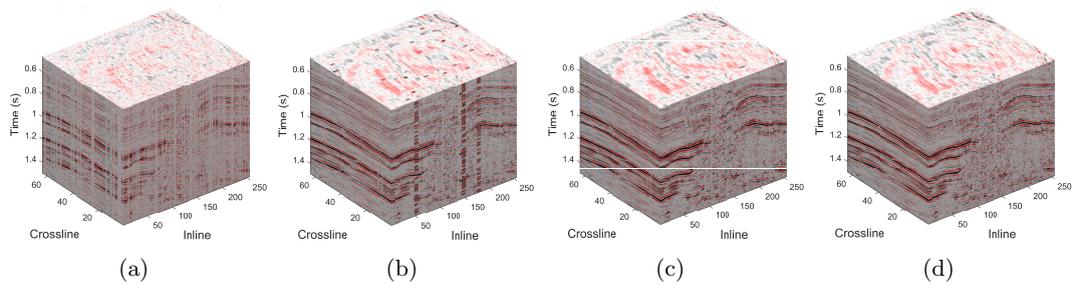

Figure 16: Input noisy data with 30% missing traces (a), the denoised and interpolated data using the OMP (b), the K-SVD (c), and the LoBCoD (d) algorithm.

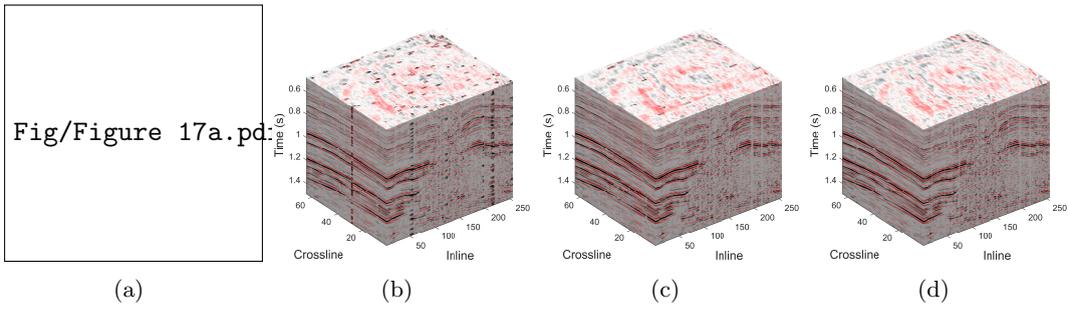

Figure 17: Input noisy data with 40% missing traces (a), the denoised and interpolated data using the OMP (b), the K-SVD (c), and the LoBCoD (d) algorithm.

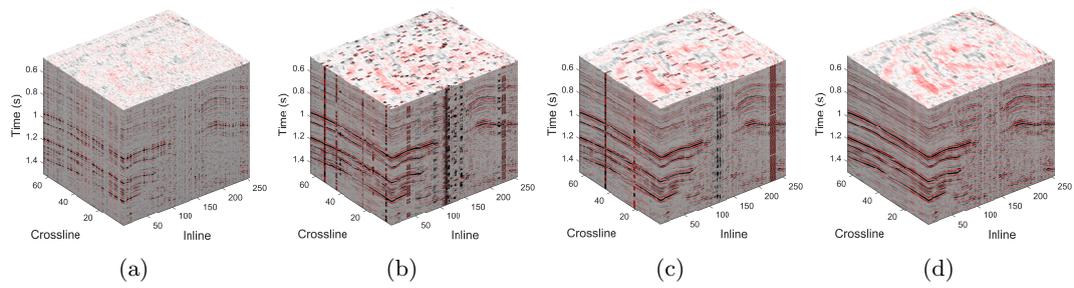

Figure 18: Input noisy data with 50% missing traces (a), the denoised and interpolated data using the OMP (b), the K-SVD (c), and the LoBCoD (d) algorithm.

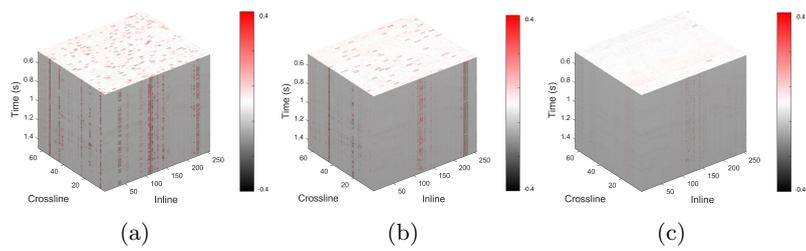

Figure 19: Interpolation errors using (a) OMP, (b) K-SVD, (c) LoBCoD algorithm for the results shown in Figure 18.

# LIST OF TABLES

Table 1: Comparison of denoising performances using the OMP, K-SVD, and LoBCoD algorithms for data with different input PSNR values.

| Input PSNR (dB) | OMP (dB) | K-SVD (dB) | LoBCoD (dB) |
|---|---|---|---|
| 20 | 32.38 | 33.43 | 34.31 |
| 15 | 29.21 | 30.34 | 31.8 |
| 10 | 25.83 | 26.32 | 28.73 |

Table 2: Comparison of denoising performances using the OMP, K-SVD, and LoBCoD algorithms for data with different input RLNE values.

| Input RLNE | OMP | K-SVD | LoBCoD |
|---|---|---|---|
| 1.48 | 0.35 | 0.32 | 0.29 |
| 2.64 | 0.51 | 0.45 | 0.38 |
| 4.7 | 0.75 | 0.72 | 0.54 |

Table 3: Comparison of denoising performances using the OMP, K-SVD, and LoBCoD algorithms for data with different input SSIM values.

| Input SSIM | OMP | K-SVD | LoBCoD |
|:---:|:---:|:---:|:---:|
| 0.1 | 0.61 | 0.58 | 0.83 |
| 0.04 | 0.43 | 0.41 | 0.78 |
| 0.02 | 0.27 | 0.25 | 0.75 |

Table 4: Average energy calculations for the reconstructed data shown in Figure 12. S1 and S2 denote the average energy of the signals before and after 2 s, respectively.

|  | Noisy Field Data | Denoised using OMP | Denoised using K-SVD | Denoised using LoBCoD |
|---|---|---|---|---|
| S1 | $0.19637 \times 10^{-3}$ | $0.18729 \times 10^{-3}$ | $0.16636 \times 10^{-3}$ | $0.16341 \times 10^{-3}$ |
| S2 | 0.0168 | 0.0168 | 0.0166 | 0.0167 |
| S2 / S1 | 85.65 | 89.53 | 99.57 | 102.06 |

Table 5: PSNR, RLNE and SSIM values of the interpolated results using the OMP, K-SVD, and LoBCoD algorithms for the data with 30% missing traces.

|           | Input data with 30% missing traces | Interpolated using OMP | Interpolated using K-SVD | Interpolated using LoBCoD |
|-----------|-----------------------------------|------------------------|--------------------------|---------------------------|
| PSNR (dB) | 20                                | 25.73                  | 29.89                    | 33.93                     |
| RLNE      | 0.15                              | 0.12                   | 0.06                     | 0.04                      |
| SSIM      | 0.46                              | 0.82                   | 0.86                     | 0.97                      |

Table 6: PSNR, RLNE and SSIM values of the interpolated results using the OMP, K-SVD, and LoBCoD algorithms for the data with 40% missing traces.

|  | Input data with 40% missing traces | Interpolated using OMP | Interpolated using K-SVD | Interpolated using LoBCoD |
|---|---|---|---|---|
| PSNR (dB) | 20 | 22.01 | 26.68 | 32.56 |
| RLNE | 0.16 | 0.15 | 0.09 | 0.05 |
| SSIM | 0.41 | 0.63 | 0.8 | 0.93 |

Table 7: PSNR, RLNE and SSIM values of the interpolated results using the OMP, K-SVD, and LoBCoD algorithms for the data with 50% missing traces.

|  | Input data with 50% missing traces | Interpolated using OMP | Interpolated using K-SVD | Interpolated using LoBCoD |
|---|---|---|---|---|
| PSNR (dB) | 20 | 21.55 | 23.49 | 29.63 |
| RLNE | 0.17 | 0.14 | 0.12 | 0.04 |
| SSIM | 0.35 | 0.41 | 0.6 | 0.91 |